\documentclass[aps,prb,superscriptaddress,twocolumn,showpacs]{revtex4-1}

\usepackage{graphicx}
\usepackage{amsmath, amsfonts, amssymb, mathrsfs}
\usepackage{xspace}
\usepackage{color}
\usepackage[colorlinks=true,linkcolor=blue,urlcolor=blue,citecolor=red]{hyperref}

\usepackage[normalem]{ulem} 

\definecolor{linkcolor}{rgb}{1,0,0}

\DeclareMathOperator{\Tr}{\mbox{Tr}}

\DeclareMathOperator{\im}{\mbox{Im}}

\newcommand{\ie}{i.e.\@\xspace}

\newcommand{\ve}[1]{{\bf #1}}
\newcommand{\mat}[1]{\mathsf{#1}}
\newcommand{\nag}{{\phantom{\dagger}}}

\newcommand{\eqw}[1]{(\ref{#1})}
\newcommand{\eq}[1]{Eq.\thinspace{}(\ref{#1})}
\newcommand{\eqq}[2]{Eqs.\thinspace{}(\ref{#1}) and (\ref{#2})}

\newcommand{\fig}[1]{Fig.\thinspace{}\ref{#1}}

\newcommand{\fc}[1]{({#1})}
\newcommand{\figc}[2]{Fig.\thinspace{}\ref{#1}\thinspace{}\fc{#2}}
\newcommand{\figcc}[3]{Fig.\thinspace{}\ref{#1}\thinspace{}\fc{#2} and \fc{#3}}

\def\ket#1{\mathinner{|{#1}\rangle}}

\begin{document}

\title{Vibration-mediated correlation effects in the transport properties of a benzene molecule}

\author{Michael Knap}
\affiliation{Institute of Theoretical and Computational Physics, Graz University of Technology, 8010 Graz, Austria}

\author{Enrico Arrigoni}
\affiliation{Institute of Theoretical and Computational Physics, Graz University of Technology, 8010 Graz, Austria}

\author{Wolfgang von der Linden}
\affiliation{Institute of Theoretical and Computational Physics, Graz University of Technology, 8010 Graz, Austria}

\date{\today}

\begin{abstract}  
We theoretically analyze correlation effects on the transport properties of a benzene
molecule that are mediated by interactions between the motion of the nuclei
and the transmitted charge. We focus on the lowest-lying molecular vibrational mode which allows us to 
derive an analytic expression for the current. The results provide transparent
interpretations of various features of the highly nonlinear current-voltage 
characteristics, which is experimentally accessible through resonant inelastic 
electron-tunneling spectroscopy. 
\end{abstract}


\maketitle

Vibrations  of atoms in molecules or solids 
play an important role for many intriguing phenomena in
various fields of physics. In condensed matter novel states often emerge from phonon mediated collective behavior. Among them are conventional superconductivity,~\cite{bardeen_theory_1957}
colossal magnetoresistance,~\cite{millis_colossal_1998} and thermoelectricity.~\cite{delaire_giant_2011} 
Localized vibrational 
modes are also believed to stabilize the $\alpha$-helix in protein and to be responsible for storage and 
transport processes.~\cite{scott_davydovs_1992} 
In molecular electronics, the electron-phonon interactions
strongly influence the conduction properties, contribute
significantly to heating, and lead to nonlinear effects, including bistability, negative 
differential conductance, and hysteretic behavior.~\cite{flensberg_tunneling_2003,koch_franck-condon_2005,galperin_resonant_2006,ryndyk_nonequilibrium_2006,benesch_vibronic_2006,ryndyk_nonequilibrium_2007,galperin_molecular_2007,galperin_nuclear_2008,botelho_unified_2011,cuniberti_2005,cuevas_molecular_2010,song_single_2011}
In these systems signatures of the interaction between charge carriers and molecular vibrations can be faithfully characterized by 
(resonant) inelastic electron tunneling [(R)IET] spectroscopy.~\cite{jaklevic_molecular_1966,stipe_single-molecule_1998,cuevas_molecular_2010} 
The importance of vibrational modes on the conduction
properties of molecular transistors has been demonstrated in various seminal experiments.~\cite{park_nanomechanical_2000,zhitenev_conductance_2002,smit_measurement_2002,qiu_vibronic_2004,
pasupathy_vibration-assisted_2005,tal_electron-vibration_2008,kiguchi_highly_2008,song_observation_2009}

A wide range of theoretical studies concerning nonequilibrium properties of nanoscale
devices are based on the nonequilibrium Green's function (NEGF) method  
evaluated in the frame of the local density approximation (LDA) 
(see e.g. Ref.~\onlinecite{frederiksen_inelastic_2007,cuniberti_green_2009} and references therein).
On the one hand, the LDA-NEGF approach is a \textit{first-principles approach}
and as such makes it possible to study realistic devices, on the other hand, correlation effects,
irrespective whether they are of electron-electron or electron-phonon character, are included 
only to low order. This can lead to unphysical results provided the interaction 
is strong.~\cite{lee_exploring_2009}

\begin{figure}
\begin{center}
 \includegraphics[width=0.48\textwidth]{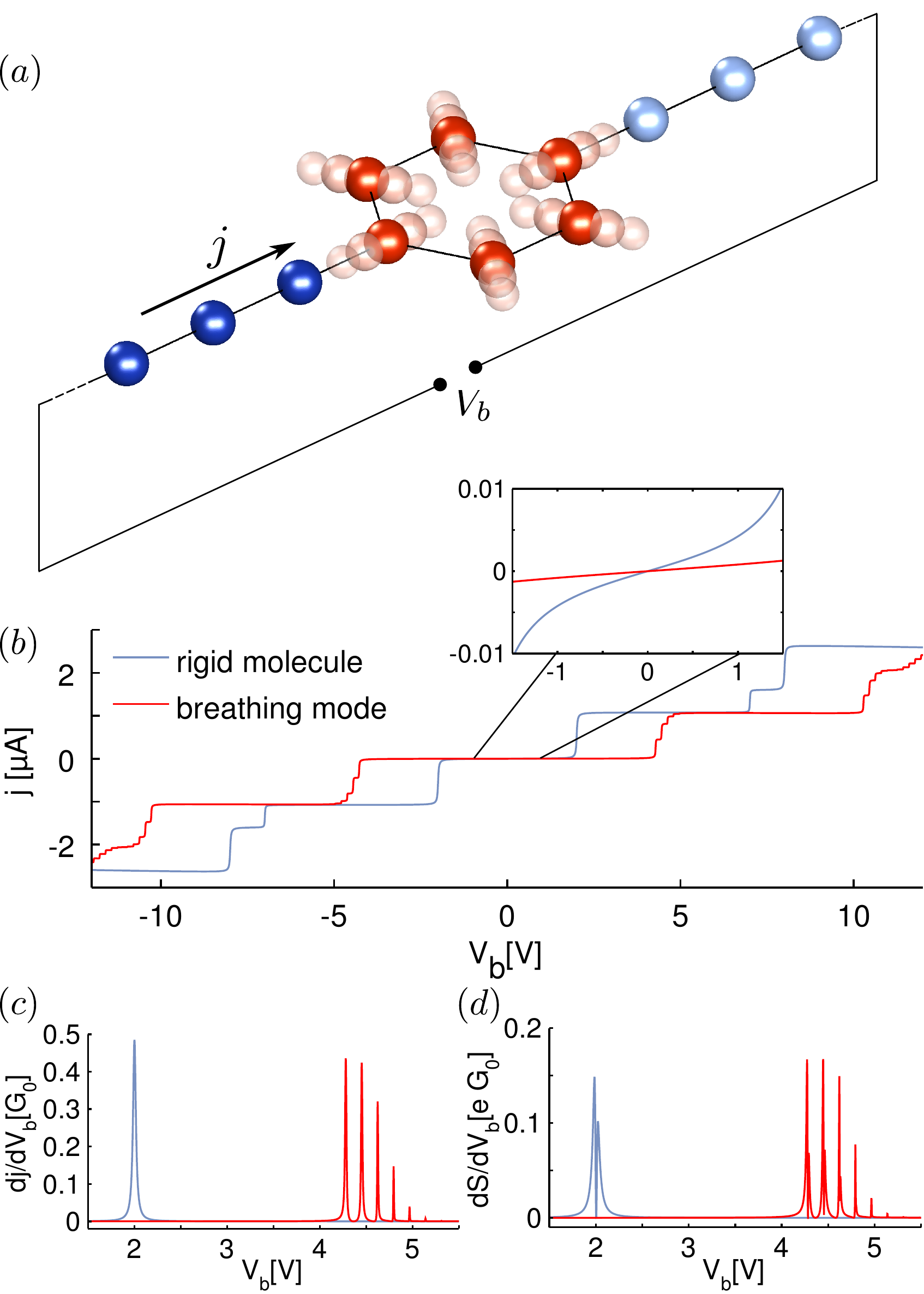}
\end{center}
\caption{\label{fig:benz} (Color online) \fc{a} Illustration of the phonon breathing mode
in a benzene molecule (red spheres) which is coupled to metallic leads (blue
spheres). \fc{b} The bias voltage $V_b$ induces a current $j$ that is significantly modified
by the interaction between the nuclear motion and the charge. At low $V_b$
the current is suppressed,~\cite{koch_franck-condon_2005} see inset, 
whereas at large $V_b$ strong nonlinear behavior leads to a stepwise increase
in the current, also manifesting in \fc{c} the conductance and \fc{d}
the differential shot noise through multiple equally spaced peaks. The data 
is evaluated at $T=10$K. }
\end{figure}

In the theoretical approach, presented here, we start out from the exact 
expression for the current within the NEGF framework and employ a strong 
coupling cluster approximation to evaluated the required Green's 
functions (GF).~\cite{knap_noneq_2011,nuss_steady-state_2012} This allows to 
exactly take into account all interaction of the nanoscale device itself, 
be it of electron-electron or electron-phonon nature, while treating the 
influence of the leads perturbatively. 
As an application, we investigate the nonequilibrium transport across 
a \textit{benzene} molecule [\figc{fig:benz}{a}] and observe 
that at low temperatures phonon mediated interactions leave a vibrational 
fingerprint qualitatively similar to those observed in (R)IET 
spectroscopy.~\cite{park_nanomechanical_2000,zhitenev_conductance_2002,smit_measurement_2002,qiu_vibronic_2004,
pasupathy_vibration-assisted_2005,tal_electron-vibration_2008,kiguchi_highly_2008,song_observation_2009}

At low temperatures we obtain a stepwise
increase in the current at voltage differences given by twice the phonon energy, \figc{fig:benz}{b}. 
This stepwise increase manifests in the conductance and in the differential shot 
noise through multiple, equally-spaced peaks, \figcc{fig:benz}{c}{d}, as well as in a branching 
of the diamond structure in the stability diagram, \fig{fig:sd}. At low bias voltage the 
current is significantly suppressed due to electron-phonon interactions, which is reminiscent 
of the Franck-Condon blockade in molecular devices,~\cite{koch_franck-condon_2005} see inset of \figc{fig:benz}{b}.

\section{Theoretical description}
We consider a 
single benzene aromatic ring coupled to metallic leads, \figc{fig:benz}{a}. 
The benzene molecule is modeled by the Hamiltonian
\begin{align}
  H_b &=   H_{e}+  H_{p} +  H_{ep}\;. \label{eq:hb}
\end{align}
The first term describes the motion of the electrons, 
\begin{align*}
 H_{e} &= \sum_{i, \sigma} \left[-t(c_{i\sigma}^\dag c_{i+1 \sigma}^\nag + \text{h.c.})  + \epsilon c^\dag_{i\sigma}c^\nag_{i\sigma} \right]\;, 
\end{align*}
where $t$ is the hopping amplitude and $\epsilon$ the local energy, which can
be controlled by the gate voltage $V_g$, \ie, $\epsilon \rightarrow \epsilon + eV_g$.
Electrons with spin $\sigma$ on orbital $i$ are created (destroyed) by $c_{i\sigma}^\dag$ 
($c_{i\sigma}^\nag$). In this paper, we consider an effective twelve spin-orbital model, describing
the delocalized $\pi$ bonds of benzene. The second term $ H_p$ is the bare vibrational part, which 
we transform into the eigenmodes
\[
  H_p = \sum_m \hbar \omega_m  b_m^\dag b^\nag_m\;,
\]
where $b_m^\dag$ ($b_m^\nag$) create (destroy) phonons in mode $m$ with 
energy $\hbar \omega_m$. The last term of \eqw{eq:hb} incorporates the 
Su-Schrieffer-Heeger~\cite{heeger_solitons_1988} interaction between the electrons 
and the dynamic phonons
\[
  H_{ep} =  \sum_{m j \sigma} g_{m} 
  (b_m^\dag + b_m)(D^{(j,j+1)}_m c_{j\sigma}^\dag c_{j+1 \sigma}^\nag + \text{h.c.})
\]
with $g_m := g \sqrt{  {\hbar}/{2 M \omega_{m}}}$, $g$ the electron-phonon
interaction, $M$ the mass of the nuclei, and $D_m$ the displacement matrix 
of eigenmode $m$, which is given by
\begin{align*}
D^{(j,j+1)}_{m}&=\frac{1}{\sqrt{6}}e^{i \frac{\pi}{3}  m j}\quad\text{with} \quad
m \in\{0,1,\ldots\,5\}\;.
\end{align*}
The metallic leads are described by semi-infinite tight-binding chains with 
hopping $t_\alpha$, on-site energy $\epsilon_\alpha$, and chemical potential $\mu_\alpha$ 
($\alpha=l$eft or $r$ight). We assume that the bias voltage $eV_b=\mu_l - \mu_r$ is 
applied symmetrically, \ie, $\mu_l=\epsilon_l=-\mu_r=-\epsilon_r$ and that the leads 
are equilibrated, \ie, that the occupation follows the Fermi-Dirac statistics 
$f_\alpha= [e^{(\varepsilon-\mu_\alpha)/k_BT}+1]^{-1}$, where $k_B$ is the
Boltzmann constant and $T$ the temperature.
\begin{figure*}
\begin{center}
 \includegraphics[width=0.98\textwidth]{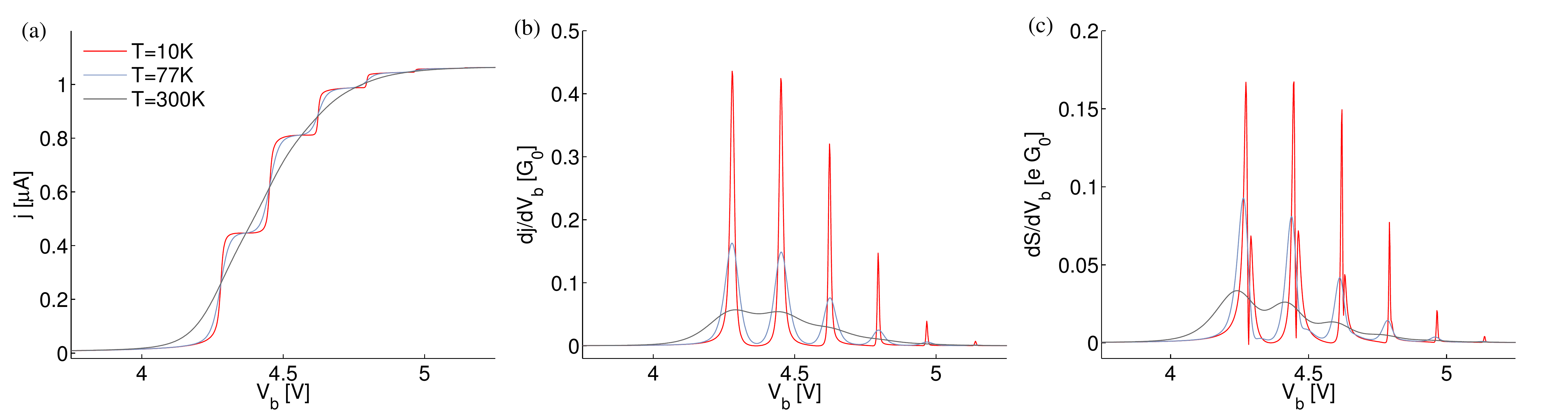}
\end{center}
\caption{\label{fig:finiteT} (Color online) Temperature dependence of \fc{a} the current, \fc{b} the conductance,
and \fc{c} the differential shot noise versus bias voltage $V_b$ curves. At temperatures $T$ much lower than the phonon energy $\omega_\mu=0.086$eV,
which is the smallest energy scale in \eq{eq:hb}, the phonon mediated steps and peaks, respectively, are pronounced and clearly
visible. For increasing temperature these features are washed out and at room temperature $T=300$K hardly any
remain. 
}
\end{figure*} 

In this paper, we use the NEGF framework to evaluate the nonequilibrium steady-state 
properties along the lines presented in Ref.~\onlinecite{knap_noneq_2011}.  Due to the electron-phonon 
interaction it is in general not possible to solve the problem exactly and thus we 
employ Cluster Perturbation Theory (CPT):~\cite{snchal_spectral_2000}
In short, the system is first divided into several clusters, each of which 
can be solved exactly either analytically or numerically, taking correlations
on the cluster level into account exactly. Strong coupling CPT is then employed to connect the individual 
clusters. 

Here, we divide the molecular device into three clusters: (i) the central region consisting of the benzene ring with vibrational degrees
of freedom and (ii) the left and (iii) right metallic leads which are connected to the molecular device.
From the exact solution of the isolated interacting central region we obtain an approximate self-energy of the full system. 
The results can also be systematically improved by incorporating lead sites into the 
central system.~\cite{knap_noneq_2011,nuss_steady-state_2012}
Based on considerations similar to those of Meir and Wingreen~\cite{me.wi.92} 
we can show that within this CPT approach the current is 
of the Landauer-B\"uttiker form~\cite{landauer_electrical_1970,buttiker_four-terminal_1986} even though correlations are 
included exactly on the cluster level. In particular, we find 
\begin{equation}
 j = \frac{e}{\hbar} \int \frac{d\varepsilon}{2\pi} ( f_l-f_r) \Tr [ \mathcal{T} ]\;,
\label{eq:j}
\end{equation}
where $\mathcal T$ is the transmission coefficient matrix
\begin{equation}
 \mathcal{T} :=
 \mathcal{G}^R(\varepsilon) \Gamma^{}_{r}(\varepsilon) \mathcal{G}^A(\varepsilon) \Gamma^{}_{l}(\varepsilon)\;.
\label{eq:jT}
\end{equation}
This matrix contains the retarded (advanced) GF $\mathcal{G}^R$ 
($\mathcal{G}^A=(\mathcal{G}^R)^\dag$) which within CPT is given by 
$(\mathcal{G}^R)^{-1}=({g}^R)^{-1}- (\tilde \Sigma_l + \tilde \Sigma_r) $,
where $g^R$ is the exact retarded GF of the isolated benzene molecule, 
$\tilde \Sigma_{\alpha}:={\mat{T}}_{c\alpha} g_{\alpha\alpha}^R {\mat{T}}_{\alpha c}$ 
takes into account the effect of lead $\alpha$ on the GF, ${\mat{T}}_{c\alpha}$
is the tunneling coupling between the central device and lead $\alpha$, and
$g_{\alpha\alpha}^R$ stands for the retarded lead GF.~\cite{knap_noneq_2011}
In addition, the transmission $\mathcal T$ depends on the imaginary part
of the lead induced self-energy via $\Gamma_\alpha = -2 \im \tilde \Sigma_\alpha$.
Similarly, the expression for the shot noise, \ie, current-current correlations,
in the zero frequency limit reads~\cite{zhu_theory_2003}
\begin{align}
 S(\omega \to 0 ) := S &= \frac{e^2}{\hbar} \int \frac{d\varepsilon}{2\pi} [ f_l(1-f_l)+f_r(1-f_r)] \Tr[\mathcal{T}] \nonumber \\
   &+ (f_l-f_r)^2 \Tr[(1-\mathcal{T})\mathcal{T}]\label{eq:s}\;.
\end{align}

In principle the nonequilibrium properties can be determined by 
solving the eigenvalue problem of $H_{b}$ numerically 
taking all vibrational modes into account using the formulas discussed above. 
Within this approach it is also straight forward to introduce electron-electron interactions giving rise to 
Kondo physics provided the molecule is (asymmetrically) electron or hole doped.~\cite{romeike_quantum_2006} 

In the present paper, however, we want to study the main effects generated by the electron-phonon interaction.
In order to gain insight into the problem we  focus on a single vibrational mode, in particular on the  
breathing mode illustrated in \figc{fig:benz}{a}. Coupling the benzene ring to the
leads breaks the C6 rotational invariance of the molecule. However, calculations reveal that
a breathing-like mode should still exists in that case. A very appealing feature of this somewhat
simplified mode is the fact that the resulting interacting model of the central region can 
be solved analytically and thus provides a transparent insight into phonon mediated 
correlation effects on the non-equilibrium physics of such a molecular electronic device.

\section{Analytic solution}
For a single vibrational mode $\mu$ (not necessarily the breathing mode)
we can readily eliminate the electron-phonon interaction by a 
unitary transformation of the Lang-Firsov type.~\cite{cuniberti_green_2009} To this end
we first diagonalize the displacement matrix $D_\mu X = X \Lambda$, $\Lambda$ is the 
diagonal matrix with the eigenvalues and $X$ the corresponding 
matrix of eigenvectors, and transform the electron operators accordingly $\ve d = X \ve c$,
where we introduce the vector notation  $\ve c^\dag = (c_{1\uparrow}^\dag,c_{2\uparrow}^\dag,\,\ldots\,c_{L\downarrow}^\dag)$.
The Lang-Firsov like unitary transformation is generated by the operator
\begin{equation}
e^S:= 
 e^{\frac{g_\mu}{\hbar \omega_\mu} 
 (b- b^\dag)
  \ve d^\dag \Lambda \ve d} \;.
\end{equation}
All operators that are subject to the unitary transformation are denoted by $\bar O:=
e^{S} O e^{-S}$, which in particular results in 
\begin{align}
\bar d_{k\sigma}^\dag&=  d_{k\sigma}^\dag e^{\frac{g_\mu}{\hbar \omega_\mu} 
 (b - b^\dag)
\lambda_k}\;.\label{eq:d}
\end{align}
The Lang-Firsov transformation eliminates the explicit electron-phonon interaction and results in an effective Hamiltonian 
$\bar H_b = \bar H_{e} + \hbar \omega_{\mu} b^{\dagger}b$.
The electronic GF  
entering the transport properties of the molecular junction [\eq{eq:jT}] is 
transformed accordingly, \ie, ${g}(k,\omega)\rightarrow \bar{g}(k,\omega)$ by replacing 
$d_{k\sigma}^\dag \rightarrow \bar d_{k\sigma}^\dag $ in its definition.
For the particular case of the breathing mode the matrices $D_{\mu}$ and the matrix
describing the nearest neighbor hopping commute and thus  can be diagonalized by a common set of eigenvectors. 
In this case we find 
$\lambda_k=\sqrt{2/3} \cos{k}$ and $X_{nk} = \exp(i k n )/\sqrt{6}$. 
The $d_k^\dag$ thus create electrons with a given (angular) momentum $k$ in
the molecule. In the transformed Hamiltonian the electronic part becomes
\begin{align}
 \bar{H}_e  &=\sum_{k \sigma} (\epsilon-2 t \cos k) \bar b^\dag_{k\sigma}\bar b^\nag_{k\sigma} 
 - U\left(\sum_{k\sigma}  \bar b^\dag_{k\sigma}\bar b^\nag_{k\sigma} \cos k \right)^2\;.\label{eq:He}
\end{align}
After this transformation, the electron and vibrational degrees of freedom are completely decoupled but instead 
an attractive electron-electron interaction proportional to  
$U:={\frac{2 g_\mu^2}{3 \hbar \omega_\mu}}$ emerges. 
At first sight Hamiltonian \eqw{eq:He} might be reminiscent of the Holstein Hamiltonian;
note however, that the parentheses enclose the sum in the second term, which gives rise 
to additional contributions mixing different angular momenta $k$. 

The interacting Hamiltonian \eqw{eq:He} can be solved exactly. The eigenvectors are of the
form $\ket{n}\ket{K_{\uparrow}}\ket{K_{\downarrow}}$, \ie, they are
tensor products of the oscillator eigenvectors $\ket{n}$ and the Slater determinants
$\ket{K_{\sigma}}:=\prod_{k\in K_{\sigma}} \bar d^{\dagger}_{k\sigma} \ket{0}$ are
composed of the molecular orbital operators $d_{k\sigma}^{\dagger}$ dressed by bond-distortions 
that are encoded in the Lang-Firsov factor of \eq{eq:d}. In \eq{eq:He} the dressed particles (polarons)
are conserved, however, the current is calculated for the electrons, which 
exhibit inelastic electron-phonon scattering. For the electron GF of a single spin-species we find
\begin{equation}
\begin{split}
  \bar{g}(k,\omega)&=\sum_n  \frac{e^{-\alpha^2_k}\alpha^{2n}_k}{n!} \\&
\left[ \frac{\Theta(k>k_F)}{\hbar \omega-(E^{+}_{kn}-E_0)}+\frac{\Theta(k<k_F)}{\hbar\omega+(E^{-}_{kn} -E_0)} \right]
  \end{split}
    \label{eq:Gex}
\end{equation}
with $\alpha_k=\sqrt{U} \cos k$ and the energies are
\begin{align*}
 E_0&=\epsilon\mathcal{N}_0 -2t \mathcal{C}_0 - U \mathcal{C}_0^2 \\
 E^{\pm}_{kn}&=n \hbar\omega_\mu  + (\mathcal{C}_0\pm1) \epsilon-2t (\mathcal{C}_0\pm\cos k) - 
U
 (\mathcal{C}_0\pm\cos k)^2 \ \\
 \mathcal{C}_0 &= {\sum_{\sigma}\sum_{q \in \text{FS}} \cos q}\;.
\end{align*}
$\mathcal{N}_0$ is the equilibrium number of electrons in the benzene molecule
and FS denotes the Fermi sea.
The physical properties of the system are particularly transparent
in the differential conductance which up to order $O(V_b^2/T^{2})$ and $O(t'^2/t_\alpha^2)$ is given by 
\begin{align}
\frac{d j}{d V_{b}} &= \frac{e}{4\hbar \pi} \left[ \mathcal T ({V_{b}}/{2})
+
\mathcal T (-{V_{b}}/{2}) 
\right] \;.\label{eq:diff:cond}
\end{align}
Along with \eqq{eq:jT}{eq:Gex} it can be deduced that current pathways open up, whenever ${V_{b}}/{2}$
is equal to an the excitation energy \hbox{$\hbar \omega =E_{kn}^{+}+\hbar \omega_{\mu} n$} in the particle sector
or \hbox{$\hbar \omega =-E_{kn}^{-}+\hbar \omega_{\mu} n$} in the hole sector.

\section{Transport properties}
Next, the impact of phonons
on the transport properties will be investigated guided by the breathing mode of a 
benzene molecule. We use the parameters reported in Ref.~\onlinecite{rai_circular_2010,botelho_unified_2011} for the
benzene molecule and the electron-phonon coupling; in particular we set $t=2.5$eV,
$\epsilon=-1.5$eV, $\hbar \omega_\mu=\sqrt{K/M}=0.086$eV, and $g_\mu=0.18$eV. In addition, for the metallic
leads we choose $t_\alpha=12$eV 
close to the wideband limit and the coupling
between the leads and the benzene ring is $t'=0.4$eV. Therefore the smallest energy scale
in the model is the phonon energy $\hbar \omega_\mu$. The current evaluated at $T=10$K  
(the temperature enters through the Fermi-Dirac distribution functions of the leads) 
for the 
rigid molecule as well as the molecule with the breathing mode is shown in 
\figc{fig:benz}{b}. At low bias voltage we observe a suppression of the current
due to electron-phonon interactions, reminiscent of the Franck-Condon blockade,~\cite{koch_franck-condon_2005}
see inset. Due to the electron-phonon mediated correlations 
(second term in \eq{eq:He})
the gap in the density
of states of the benzene molecule widens considerably. Consequently, 
the threshold for the current increases to $V_b\sim4$eV. Above the threshold the 
current increases stepwise, at voltage differences given by twice 
the phonon energy $\hbar \omega_\mu$, until the next
plateau is reached. The new current pathways are related to the vibrational 
excitations and the factor 2 is due to the argument $V_{b}/2$ in \eq{eq:diff:cond}. 

This also manifests in the differential conductance \figc{fig:benz}{c} and in the
differential shot-noise \fc{d} through pronounced, equally spaced peaks. At the maxima
of the conductance, the differential shot-noise, corresponding to the derivative of
the current-current correlation for $\omega\to0$, exhibits dips.~\cite{zhu_theory_2003} 
At low temperatures this is observable for a few low-energy peaks of the phonon branching.
A detailed analysis of the differential conductance shows that the peaks are
near the poles of the cluster GF $\bar g$ of the decoupled device. This is reasonable, since the effective coupling $V/t'^2$ is small. Therefore, the impact of $\tilde \Sigma_\alpha$ stems primarily from its imaginary part which determines the width
and the weight of the individual peaks.
So the measurement of these quantities can be used as a diagnostic tool to analyze the device and to extract information about the electronic and phononic excitations. In particular the equidistant peaks in 
the conductance and the tips in the differential shot-noise are a fingerprint of the bare vibrational excitations.
One should notice, however, that for general vibrational modes the distance between the peaks does not need to be equidistant, since the electronic excitations will depend on the phonon occupation $n$, due to $\bar H_{e}$. 
\begin{figure}
\begin{center}
 \includegraphics[width=0.48\textwidth]{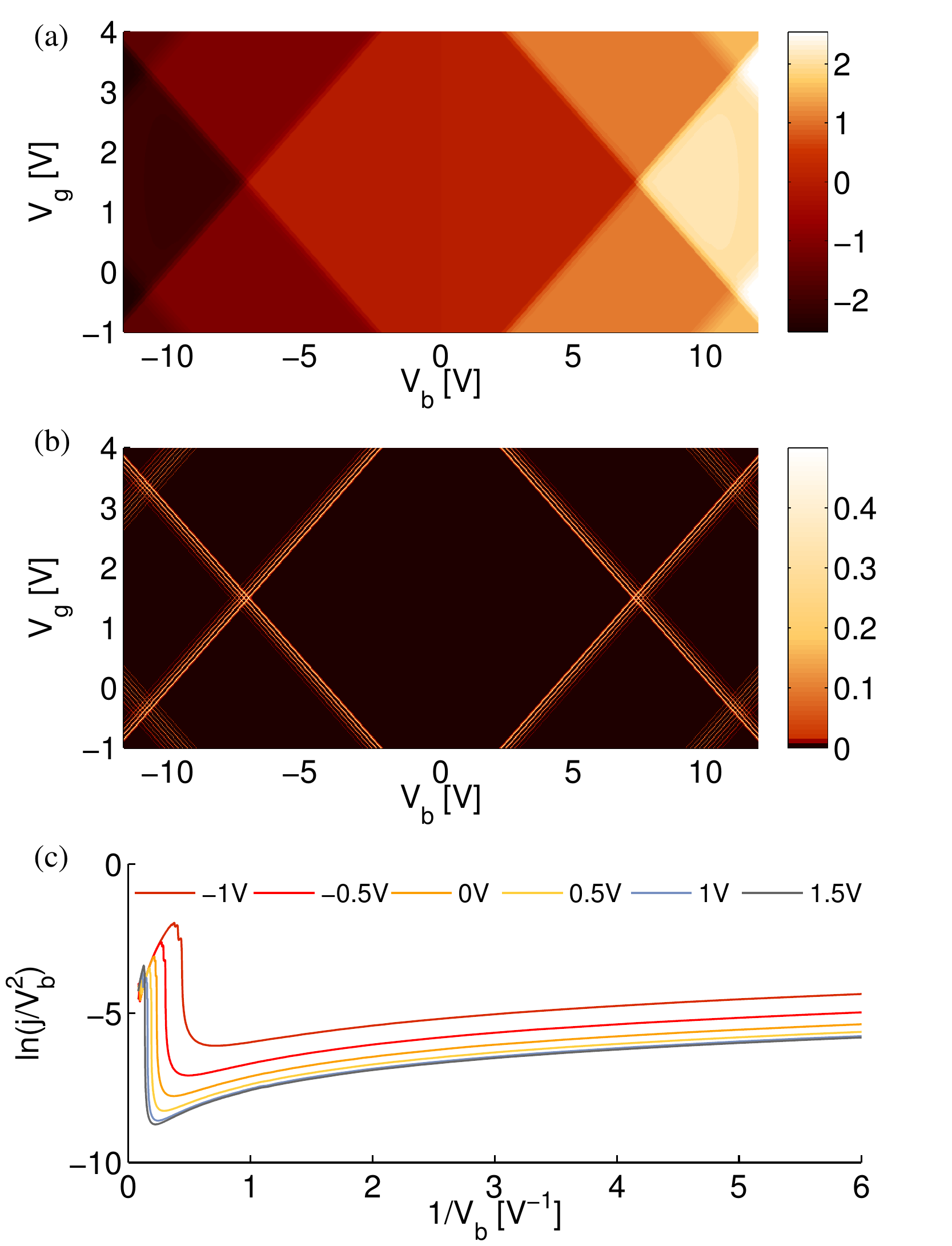}
\end{center}
\caption{\label{fig:sd} (Color online) \fc{a} Current as a function of the bias voltage $V_b$
and the gate voltage $V_g$ and \fc{b} corresponding stability diagram (\ie, differential 
conductance). The electron-phonon interaction increases the area of the central diamond and induces
a splitting of the lines in the stability diagram. \fc{c} Fowler-Nordheim plots for a few
selected values of the gate voltage $V_g$ specified in the legend. Data evaluated for $T=10$K.
}
\end{figure}

For increasing temperature the vibrational features are smeared out,
since the phonon energy $\omega_\mu=86$meV is rather small and comparable to room temperature
$T=300\text{K}=26$meV. In \fig{fig:finiteT}, the current, the differential conductance, and the differential
shot noise are shown at different temperatures $T=\lbrace 10,\,77,\,300\rbrace$K. With increasing
temperature the steps in the current are smoothed out and the peaks in the conductance and the differential 
shot noise decrease significantly. In addition, the dips in the shot noise vanish.

Finally, we study the consequence of orbital gating, where the local on-site energy
of the benzene molecule is changed by a gate voltage. Experimentally orbital gating has recently been
achieved in Ref.~\onlinecite{song_observation_2009}.
The current as a function of the gate voltage $V_g$ and
the bias voltage $V_b$ exhibits the celebrated diamond structure centered around the
particle-hole symmetric point $V_g=1.5$V, \figc{fig:sd}{a}.
As compared to the rigid molecule, see \figc{fig:benz}{b}--\fc{d}, the area of the center diamond is increased due to the
electron-phonon interactions and the transition between the diamonds involve several steps,
best visible in the stability diagram, corresponding to the conductance in the $V_g$-$V_b$
plane, shown in \figc{fig:sd}{b}. The Fowler-Nordheim plots \fc{c} which show $\ln (j/V_b^2)$ 
as a function of $V_b^{-1}$ compare well with the experimental observation in 
Ref.~\onlinecite{song_observation_2009}. From these plots different tunneling regimes become apparent: 
At low bias voltage, the ``logarithmic'' dependence indicates direct tunneling, whereas
for increasing bias voltage the curve has a large negative slope, indicating
field emission.~\cite{song_observation_2009} Close to this bias voltage phonon mediated interactions manifest as 
small wiggles in the curves. When further increasing the voltage the next plateau is reached 
which is accompanied by a reentrance in the direct tunneling regime.  Similarly
to Ref.~\onlinecite{song_observation_2009} the minimum of the curves move to larger bias voltage for increasing 
gate voltage reflecting the diamond shape of \figcc{fig:sd}{a}{b}.

\section{Conclusions and Outlook}
To summarize, we explored phonon mediated correlation effects on the 
transport properties of a single benzene molecule. We focused on
an analytically tractable single-mode model that reveals a deeper 
understanding of the impact of phonons on experimental (R)IET spectra. 
The theoretical approach we presented makes it possible to determine non-equilibrium
properties of strongly correlated nano-scale devices. When solving 
the interacting device totally numerically,
other vibrational modes and electron-electron interactions can be included as well.

\begin{acknowledgments}
We are grateful to G. Cuniberti, R. Guti{\'e}rrez, and D. A. Ryndyk for
insightful discussions. The authors acknowledge support from the 
Austrian Science Fund (FWF) under Project No. P24081-N16 and No. J 3361-N20. 
Numerical calculations have been performed on the Vienna Scientific 
Cluster (VSC II).
\end{acknowledgments}


\begin{thebibliography}{40}%
\makeatletter
\providecommand \@ifxundefined [1]{%
 \@ifx{#1\undefined}
}%
\providecommand \@ifnum [1]{%
 \ifnum #1\expandafter \@firstoftwo
 \else \expandafter \@secondoftwo
 \fi
}%
\providecommand \@ifx [1]{%
 \ifx #1\expandafter \@firstoftwo
 \else \expandafter \@secondoftwo
 \fi
}%
\providecommand \natexlab [1]{#1}%
\providecommand \enquote  [1]{``#1''}%
\providecommand \bibnamefont  [1]{#1}%
\providecommand \bibfnamefont [1]{#1}%
\providecommand \citenamefont [1]{#1}%
\providecommand \href@noop [0]{\@secondoftwo}%
\providecommand \href [0]{\begingroup \@sanitize@url \@href}%
\providecommand \@href[1]{\@@startlink{#1}\@@href}%
\providecommand \@@href[1]{\endgroup#1\@@endlink}%
\providecommand \@sanitize@url [0]{\catcode `\\12\catcode `\$12\catcode
  `\&12\catcode `\#12\catcode `\^12\catcode `\_12\catcode `\%12\relax}%
\providecommand \@@startlink[1]{}%
\providecommand \@@endlink[0]{}%
\providecommand \url  [0]{\begingroup\@sanitize@url \@url }%
\providecommand \@url [1]{\endgroup\@href {#1}{\urlprefix }}%
\providecommand \urlprefix  [0]{URL }%
\providecommand \Eprint [0]{\href }%
\providecommand \doibase [0]{http://dx.doi.org/}%
\providecommand \selectlanguage [0]{\@gobble}%
\providecommand \bibinfo  [0]{\@secondoftwo}%
\providecommand \bibfield  [0]{\@secondoftwo}%
\providecommand \translation [1]{[#1]}%
\providecommand \BibitemOpen [0]{}%
\providecommand \bibitemStop [0]{}%
\providecommand \bibitemNoStop [0]{.\EOS\space}%
\providecommand \EOS [0]{\spacefactor3000\relax}%
\providecommand \BibitemShut  [1]{\csname bibitem#1\endcsname}%
\let\auto@bib@innerbib\@empty
\bibitem [{\citenamefont {Bardeen}\ \emph {et~al.}(1957)\citenamefont
  {Bardeen}, \citenamefont {Cooper},\ and\ \citenamefont
  {Schrieffer}}]{bardeen_theory_1957}%
  \BibitemOpen
  \bibfield  {author} {\bibinfo {author} {\bibfnamefont {J.}~\bibnamefont
  {Bardeen}}, \bibinfo {author} {\bibfnamefont {L.~N.}\ \bibnamefont {Cooper}},
  \ and\ \bibinfo {author} {\bibfnamefont {J.~R.}\ \bibnamefont {Schrieffer}},\
  }\href {\doibase 10.1103/PhysRev.108.1175} {\bibfield  {journal} {\bibinfo
  {journal} {Phys. Rev.}\ }\textbf {\bibinfo {volume} {108}},\ \bibinfo {pages}
  {1175} (\bibinfo {year} {1957})}\BibitemShut {NoStop}%
\bibitem [{\citenamefont {Millis}(1998)}]{millis_colossal_1998}%
  \BibitemOpen
  \bibfield  {author} {\bibinfo {author} {\bibfnamefont {A.~J.}\ \bibnamefont
  {Millis}},\ }\href {\doibase 10.1098/rsta.1998.0230} {\bibfield  {journal}
  {\bibinfo  {journal} {Phil. Trans. R. Soc. London, Ser. A}\ }\textbf {\bibinfo
  {volume} {356}},\ \bibinfo {pages} {1473} (\bibinfo {year}
  {1998})}\BibitemShut {NoStop}%
\bibitem [{\citenamefont {Delaire}\ \emph {et~al.}(2011)\citenamefont
  {Delaire}, \citenamefont {Ma}, \citenamefont {Marty}, \citenamefont {May},
  \citenamefont {{McGuire}}, \citenamefont {Du}, \citenamefont {Singh},
  \citenamefont {Podlesnyak}, \citenamefont {Ehlers}, \citenamefont {Lumsden},\
  and\ \citenamefont {Sales}}]{delaire_giant_2011}%
  \BibitemOpen
  \bibfield  {author} {\bibinfo {author} {\bibfnamefont {O.}~\bibnamefont
  {Delaire}}, \bibinfo {author} {\bibfnamefont {J.}~\bibnamefont {Ma}},
  \bibinfo {author} {\bibfnamefont {K.}~\bibnamefont {Marty}}, \bibinfo
  {author} {\bibfnamefont {A.~F.}\ \bibnamefont {May}}, \bibinfo {author}
  {\bibfnamefont {M.~A.}\ \bibnamefont {{McGuire}}}, \bibinfo {author}
  {\bibfnamefont {M.-H.}\ \bibnamefont {Du}}, \bibinfo {author} {\bibfnamefont
  {D.~J.}\ \bibnamefont {Singh}}, \bibinfo {author} {\bibfnamefont
  {A.}~\bibnamefont {Podlesnyak}}, \bibinfo {author} {\bibfnamefont
  {G.}~\bibnamefont {Ehlers}}, \bibinfo {author} {\bibfnamefont {M.~D.}\
  \bibnamefont {Lumsden}}, \ and\ \bibinfo {author} {\bibfnamefont {B.~C.}\
  \bibnamefont {Sales}},\ }\href {\doibase 10.1038/nmat3035} {\bibfield
  {journal} {\bibinfo  {journal} {Nat. Mater.}\ }\textbf {\bibinfo {volume}
  {10}},\ \bibinfo {pages} {614} (\bibinfo {year} {2011})}\BibitemShut
  {NoStop}%
\bibitem [{\citenamefont {Scott}(1992)}]{scott_davydovs_1992}%
  \BibitemOpen
  \bibfield  {author} {\bibinfo {author} {\bibfnamefont {A.}~\bibnamefont
  {Scott}},\ }\href {\doibase 10.1016/0370-1573(92)90093-F} {\bibfield
  {journal} {\bibinfo  {journal} {Phys. Rep.}\ }\textbf {\bibinfo {volume}
  {217}},\ \bibinfo {pages} {1} (\bibinfo {year} {1992})}\BibitemShut {NoStop}%
\bibitem [{\citenamefont {Flensberg}(2003)}]{flensberg_tunneling_2003}%
  \BibitemOpen
  \bibfield  {author} {\bibinfo {author} {\bibfnamefont {K.}~\bibnamefont
  {Flensberg}},\ }\href {\doibase 10.1103/PhysRevB.68.205323} {\bibfield
  {journal} {\bibinfo  {journal} {Phys. Rev. B}\ }\textbf {\bibinfo {volume}
  {68}},\ \bibinfo {pages} {205323} (\bibinfo {year} {2003})}\BibitemShut
  {NoStop}%
\bibitem [{\citenamefont {Koch}\ and\ \citenamefont {von
  Oppen}(2005)}]{koch_franck-condon_2005}%
  \BibitemOpen
  \bibfield  {author} {\bibinfo {author} {\bibfnamefont {J.}~\bibnamefont
  {Koch}}\ and\ \bibinfo {author} {\bibfnamefont {F.}~\bibnamefont {von
  Oppen}},\ }\href {\doibase 10.1103/PhysRevLett.94.206804} {\bibfield
  {journal} {\bibinfo  {journal} {Phys. Rev. Lett.}\ }\textbf {\bibinfo
  {volume} {94}},\ \bibinfo {pages} {206804} (\bibinfo {year}
  {2005})}\BibitemShut {NoStop}%
\bibitem [{\citenamefont {Galperin}\ \emph {et~al.}(2006)\citenamefont
  {Galperin}, \citenamefont {Nitzan},\ and\ \citenamefont
  {Ratner}}]{galperin_resonant_2006}%
  \BibitemOpen
  \bibfield  {author} {\bibinfo {author} {\bibfnamefont {M.}~\bibnamefont
  {Galperin}}, \bibinfo {author} {\bibfnamefont {A.}~\bibnamefont {Nitzan}}, \
  and\ \bibinfo {author} {\bibfnamefont {M.~A.}\ \bibnamefont {Ratner}},\
  }\href {\doibase 10.1103/PhysRevB.73.045314} {\bibfield  {journal} {\bibinfo
  {journal} {Phys. Rev. B}\ }\textbf {\bibinfo {volume} {73}},\ \bibinfo
  {pages} {045314} (\bibinfo {year} {2006})}\BibitemShut {NoStop}%
\bibitem [{\citenamefont {Ryndyk}\ \emph {et~al.}(2006)\citenamefont {Ryndyk},
  \citenamefont {Hartung},\ and\ \citenamefont
  {Cuniberti}}]{ryndyk_nonequilibrium_2006}%
  \BibitemOpen
  \bibfield  {author} {\bibinfo {author} {\bibfnamefont {D.~A.}\ \bibnamefont
  {Ryndyk}}, \bibinfo {author} {\bibfnamefont {M.}~\bibnamefont {Hartung}}, \
  and\ \bibinfo {author} {\bibfnamefont {G.}~\bibnamefont {Cuniberti}},\ }\href
  {\doibase 10.1103/PhysRevB.73.045420} {\bibfield  {journal} {\bibinfo
  {journal} {Phys. Rev. B}\ }\textbf {\bibinfo {volume} {73}},\ \bibinfo
  {pages} {045420} (\bibinfo {year} {2006})}\BibitemShut {NoStop}%
\bibitem [{\citenamefont {Benesch}\ \emph {et~al.}(2006)\citenamefont
  {Benesch}, \citenamefont {{\v C}{\' i}{\v z}ek}, \citenamefont {Thoss},\ and\
  \citenamefont {Domcke}}]{benesch_vibronic_2006}%
  \BibitemOpen
  \bibfield  {author} {\bibinfo {author} {\bibfnamefont {C.}~\bibnamefont
  {Benesch}}, \bibinfo {author} {\bibfnamefont {M.}~\bibnamefont {{\v C}{\'
  i}{\v z}ek}}, \bibinfo {author} {\bibfnamefont {M.}~\bibnamefont {Thoss}}, \
  and\ \bibinfo {author} {\bibfnamefont {W.}~\bibnamefont {Domcke}},\ }\href
  {\doibase 10.1016/j.cplett.2006.09.003} {\bibfield  {journal} {\bibinfo
  {journal} {Chem. Phys. Lett.}\ }\textbf {\bibinfo {volume} {430}},\ \bibinfo
  {pages} {355} (\bibinfo {year} {2006})}\BibitemShut {NoStop}%
\bibitem [{\citenamefont {Ryndyk}\ and\ \citenamefont
  {Cuniberti}(2007)}]{ryndyk_nonequilibrium_2007}%
  \BibitemOpen
  \bibfield  {author} {\bibinfo {author} {\bibfnamefont {D.~A.}\ \bibnamefont
  {Ryndyk}}\ and\ \bibinfo {author} {\bibfnamefont {G.}~\bibnamefont
  {Cuniberti}},\ }\href {\doibase 10.1103/PhysRevB.76.155430} {\bibfield
  {journal} {\bibinfo  {journal} {Phys. Rev. B}\ }\textbf {\bibinfo {volume}
  {76}},\ \bibinfo {pages} {155430} (\bibinfo {year} {2007})}\BibitemShut
  {NoStop}%
\bibitem [{\citenamefont {Galperin}\ \emph {et~al.}(2007)\citenamefont
  {Galperin}, \citenamefont {Ratner},\ and\ \citenamefont
  {Nitzan}}]{galperin_molecular_2007}%
  \BibitemOpen
  \bibfield  {author} {\bibinfo {author} {\bibfnamefont {M.}~\bibnamefont
  {Galperin}}, \bibinfo {author} {\bibfnamefont {M.~A.}\ \bibnamefont
  {Ratner}}, \ and\ \bibinfo {author} {\bibfnamefont {A.}~\bibnamefont
  {Nitzan}},\ }\href {\doibase 10.1088/0953-8984/19/10/103201} {\bibfield
  {journal} {\bibinfo  {journal} {J. Phys.: Condens. Matter}\ }\textbf
  {\bibinfo {volume} {19}},\ \bibinfo {pages} {103201} (\bibinfo {year}
  {2007})}\BibitemShut {NoStop}%
\bibitem [{\citenamefont {Galperin}\ \emph {et~al.}(2008)\citenamefont
  {Galperin}, \citenamefont {Ratner}, \citenamefont {Nitzan},\ and\
  \citenamefont {Troisi}}]{galperin_nuclear_2008}%
  \BibitemOpen
  \bibfield  {author} {\bibinfo {author} {\bibfnamefont {M.}~\bibnamefont
  {Galperin}}, \bibinfo {author} {\bibfnamefont {M.~A.}\ \bibnamefont
  {Ratner}}, \bibinfo {author} {\bibfnamefont {A.}~\bibnamefont {Nitzan}}, \
  and\ \bibinfo {author} {\bibfnamefont {A.}~\bibnamefont {Troisi}},\ }\href
  {\doibase 10.1126/science.1146556} {\bibfield  {journal} {\bibinfo  {journal}
  {Science}\ }\textbf {\bibinfo {volume} {319}},\ \bibinfo {pages} {1056}
  (\bibinfo {year} {2008})}\BibitemShut {NoStop}%
\bibitem [{\citenamefont {Botelho}\ \emph {et~al.}(2011)\citenamefont
  {Botelho}, \citenamefont {Shin}, \citenamefont {Li}, \citenamefont {Jiang},\
  and\ \citenamefont {Lin}}]{botelho_unified_2011}%
  \BibitemOpen
  \bibfield  {author} {\bibinfo {author} {\bibfnamefont {A.~L.}\ \bibnamefont
  {Botelho}}, \bibinfo {author} {\bibfnamefont {Y.}~\bibnamefont {Shin}},
  \bibinfo {author} {\bibfnamefont {M.}~\bibnamefont {Li}}, \bibinfo {author}
  {\bibfnamefont {L.}~\bibnamefont {Jiang}}, \ and\ \bibinfo {author}
  {\bibfnamefont {X.}~\bibnamefont {Lin}},\ }\href {\doibase
  10.1088/0953-8984/23/45/455501} {\bibfield  {journal} {\bibinfo  {journal}
  {J. Phys.: Condens. Matter}\ }\textbf {\bibinfo {volume} {23}},\ \bibinfo
  {pages} {455501} (\bibinfo {year} {2011})}\BibitemShut {NoStop}%
\bibitem [{\citenamefont {Cuniberti}\ \emph {et~al.}(2005)\citenamefont
  {Cuniberti}, \citenamefont {Richter},\ and\ \citenamefont
  {Fagas}}]{cuniberti_2005}%
  \BibitemOpen
  \bibfield  {author} {\bibinfo {author} {\bibfnamefont {G.}~\bibnamefont
  {Cuniberti}}, \bibinfo {author} {\bibfnamefont {K.}~\bibnamefont {Richter}},
  \ and\ \bibinfo {author} {\bibfnamefont {G.}~\bibnamefont {Fagas}},\
  }\href@noop {} {\emph {\bibinfo {title} {Introducing Molecular
  Electronics}}},\ \bibinfo {edition} {1st}\ ed.\ (\bibinfo  {publisher}
  {Springer},\ \bibinfo {address} {Heidelberg, Germany},\ \bibinfo {year}
  {2005})\BibitemShut {NoStop}%
\bibitem [{\citenamefont {Cuevas}\ and\ \citenamefont
  {Scheer}(2010)}]{cuevas_molecular_2010}%
  \BibitemOpen
  \bibfield  {author} {\bibinfo {author} {\bibfnamefont {J.~C.}\ \bibnamefont
  {Cuevas}}\ and\ \bibinfo {author} {\bibfnamefont {E.}~\bibnamefont
  {Scheer}},\ }\href@noop {} {\emph {\bibinfo {title} {Molecular Electronics:
  An Introduction to Theory and Experiment {(Nanotechnology} and
  Nanoscience)}}},\ \bibinfo {edition} {1st}\ ed.\ (\bibinfo  {publisher}
  {World Scientific Publishing Company},\ \bibinfo {address} {Singapore},\
  \bibinfo {year} {2010})\BibitemShut {NoStop}%
\bibitem [{\citenamefont {Song}\ \emph {et~al.}(2011)\citenamefont {Song},
  \citenamefont {Reed},\ and\ \citenamefont {Lee}}]{song_single_2011}%
  \BibitemOpen
  \bibfield  {author} {\bibinfo {author} {\bibfnamefont {H.}~\bibnamefont
  {Song}}, \bibinfo {author} {\bibfnamefont {M.~A.}\ \bibnamefont {Reed}}, \
  and\ \bibinfo {author} {\bibfnamefont {T.}~\bibnamefont {Lee}},\ }\href
  {\doibase 10.1002/adma.201004291} {\bibfield  {journal} {\bibinfo  {journal}
  {Advanced Materials}\ }\textbf {\bibinfo {volume} {23}},\ \bibinfo {pages}
  {1583} (\bibinfo {year} {2011})}\BibitemShut {NoStop}%
\bibitem [{\citenamefont {Jaklevic}\ and\ \citenamefont
  {Lambe}(1966)}]{jaklevic_molecular_1966}%
  \BibitemOpen
  \bibfield  {author} {\bibinfo {author} {\bibfnamefont {R.~C.}\ \bibnamefont
  {Jaklevic}}\ and\ \bibinfo {author} {\bibfnamefont {J.}~\bibnamefont
  {Lambe}},\ }\href {\doibase 10.1103/PhysRevLett.17.1139} {\bibfield
  {journal} {\bibinfo  {journal} {Phys. Rev. Lett.}\ }\textbf {\bibinfo
  {volume} {17}},\ \bibinfo {pages} {1139} (\bibinfo {year}
  {1966})}\BibitemShut {NoStop}%
\bibitem [{\citenamefont {Stipe}\ \emph {et~al.}(1998)\citenamefont {Stipe},
  \citenamefont {Rezaei},\ and\ \citenamefont
  {Ho}}]{stipe_single-molecule_1998}%
  \BibitemOpen
  \bibfield  {author} {\bibinfo {author} {\bibfnamefont {B.~C.}\ \bibnamefont
  {Stipe}}, \bibinfo {author} {\bibfnamefont {M.~A.}\ \bibnamefont {Rezaei}}, \
  and\ \bibinfo {author} {\bibfnamefont {W.}~\bibnamefont {Ho}},\ }\href
  {\doibase 10.1126/science.280.5370.1732} {\bibfield  {journal} {\bibinfo
  {journal} {Science}\ }\textbf {\bibinfo {volume} {280}},\ \bibinfo {pages}
  {1732} (\bibinfo {year} {1998})}\BibitemShut {NoStop}%
\bibitem [{\citenamefont {Park}\ \emph {et~al.}(2000)\citenamefont {Park},
  \citenamefont {Park}, \citenamefont {Lim}, \citenamefont {Anderson},
  \citenamefont {Alivisatos},\ and\ \citenamefont
  {{McEuen}}}]{park_nanomechanical_2000}%
  \BibitemOpen
  \bibfield  {author} {\bibinfo {author} {\bibfnamefont {H.}~\bibnamefont
  {Park}}, \bibinfo {author} {\bibfnamefont {J.}~\bibnamefont {Park}}, \bibinfo
  {author} {\bibfnamefont {A.~K.~L.}\ \bibnamefont {Lim}}, \bibinfo {author}
  {\bibfnamefont {E.~H.}\ \bibnamefont {Anderson}}, \bibinfo {author}
  {\bibfnamefont {A.~P.}\ \bibnamefont {Alivisatos}}, \ and\ \bibinfo {author}
  {\bibfnamefont {P.~L.}\ \bibnamefont {{McEuen}}},\ }\href {\doibase
  10.1038/35024031} {\bibfield  {journal} {\bibinfo  {journal} {Nature (London)}\
  }\textbf {\bibinfo {volume} {407}},\ \bibinfo {pages} {57} (\bibinfo {year}
  {2000})}\BibitemShut {NoStop}%
\bibitem [{\citenamefont {Zhitenev}\ \emph {et~al.}(2002)\citenamefont
  {Zhitenev}, \citenamefont {Meng},\ and\ \citenamefont
  {Bao}}]{zhitenev_conductance_2002}%
  \BibitemOpen
  \bibfield  {author} {\bibinfo {author} {\bibfnamefont {N. B.}~\bibnamefont
  {Zhitenev}}, \bibinfo {author} {\bibfnamefont {H.}~\bibnamefont {Meng}}, \
  and\ \bibinfo {author} {\bibfnamefont {Z.}~\bibnamefont {Bao}},\ }\href
  {\doibase 10.1103/PhysRevLett.88.226801} {\bibfield  {journal} {\bibinfo
  {journal} {Phys. Rev. Lett.}\ }\textbf {\bibinfo {volume} {88}},\ \bibinfo
  {pages} {226801} (\bibinfo {year} {2002})}\BibitemShut {NoStop}%
\bibitem [{\citenamefont {Smit}\ \emph {et~al.}(2002)\citenamefont {Smit},
  \citenamefont {Noat}, \citenamefont {Untiedt}, \citenamefont {Lang},
  \citenamefont {Hemert},\ and\ \citenamefont
  {Ruitenbeek}}]{smit_measurement_2002}%
  \BibitemOpen
  \bibfield  {author} {\bibinfo {author} {\bibfnamefont {R.~H.~M.}\
  \bibnamefont {Smit}}, \bibinfo {author} {\bibfnamefont {Y.}~\bibnamefont
  {Noat}}, \bibinfo {author} {\bibfnamefont {C.}~\bibnamefont {Untiedt}},
  \bibinfo {author} {\bibfnamefont {N.~D.}\ \bibnamefont {Lang}}, \bibinfo
  {author} {\bibfnamefont {M.~C.}\ \bibnamefont {van Hemert}}, \ and\ \bibinfo
  {author} {\bibfnamefont {J.~M.}\ \bibnamefont {van Ruitenbeek}},\ }\href
  {\doibase 10.1038/nature01103} {\bibfield  {journal} {\bibinfo  {journal}
  {Nature (London)}\ }\textbf {\bibinfo {volume} {419}},\ \bibinfo {pages} {906}
  (\bibinfo {year} {2002})}\BibitemShut {NoStop}%
\bibitem [{\citenamefont {Qiu}\ \emph {et~al.}(2004)\citenamefont {Qiu},
  \citenamefont {Nazin},\ and\ \citenamefont {Ho}}]{qiu_vibronic_2004}%
  \BibitemOpen
  \bibfield  {author} {\bibinfo {author} {\bibfnamefont {X.~H.}\ \bibnamefont
  {Qiu}}, \bibinfo {author} {\bibfnamefont {G.~V.}\ \bibnamefont {Nazin}}, \
  and\ \bibinfo {author} {\bibfnamefont {W.}~\bibnamefont {Ho}},\ }\href
  {\doibase 10.1103/PhysRevLett.92.206102} {\bibfield  {journal} {\bibinfo
  {journal} {Phys. Rev. Lett.}\ }\textbf {\bibinfo {volume} {92}},\ \bibinfo
  {pages} {206102} (\bibinfo {year} {2004})}\BibitemShut {NoStop}%
\bibitem [{\citenamefont {Pasupathy}\ \emph {et~al.}(2005)\citenamefont
  {Pasupathy}, \citenamefont {Park}, \citenamefont {Chang}, \citenamefont
  {Soldatov}, \citenamefont {Lebedkin}, \citenamefont {Bialczak}, \citenamefont
  {Grose}, \citenamefont {Donev}, \citenamefont {Sethna}, \citenamefont
  {Ralph},\ and\ \citenamefont {{McEuen}}}]{pasupathy_vibration-assisted_2005}%
  \BibitemOpen
  \bibfield  {author} {\bibinfo {author} {\bibfnamefont {A.~N.}\ \bibnamefont
  {Pasupathy}}, \bibinfo {author} {\bibfnamefont {J.}~\bibnamefont {Park}},
  \bibinfo {author} {\bibfnamefont {C.}~\bibnamefont {Chang}}, \bibinfo
  {author} {\bibfnamefont {A.~V.}\ \bibnamefont {Soldatov}}, \bibinfo {author}
  {\bibfnamefont {S.}~\bibnamefont {Lebedkin}}, \bibinfo {author}
  {\bibfnamefont {R.~C.}\ \bibnamefont {Bialczak}}, \bibinfo {author}
  {\bibfnamefont {J.~E.}\ \bibnamefont {Grose}}, \bibinfo {author}
  {\bibfnamefont {L.~A.~K.}\ \bibnamefont {Donev}}, \bibinfo {author}
  {\bibfnamefont {J.~P.}\ \bibnamefont {Sethna}}, \bibinfo {author}
  {\bibfnamefont {D.~C.}\ \bibnamefont {Ralph}}, \ and\ \bibinfo {author}
  {\bibfnamefont {P.~L.}\ \bibnamefont {{McEuen}}},\ }\href {\doibase
  10.1021/nl048619c} {\bibfield  {journal} {\bibinfo  {journal} {Nano Lett.}\
  }\textbf {\bibinfo {volume} {5}},\ \bibinfo {pages} {203} (\bibinfo {year}
  {2005})}\BibitemShut {NoStop}%
\bibitem [{\citenamefont {Tal}\ \emph {et~al.}(2008)\citenamefont {Tal},
  \citenamefont {Krieger}, \citenamefont {Leerink},\ and\ \citenamefont {van
  Ruitenbeek}}]{tal_electron-vibration_2008}%
  \BibitemOpen
  \bibfield  {author} {\bibinfo {author} {\bibfnamefont {O.}~\bibnamefont
  {Tal}}, \bibinfo {author} {\bibfnamefont {M.}~\bibnamefont {Krieger}},
  \bibinfo {author} {\bibfnamefont {B.}~\bibnamefont {Leerink}}, \ and\
  \bibinfo {author} {\bibfnamefont {J.~M.}\ \bibnamefont {van Ruitenbeek}},\
  }\href {\doibase 10.1103/PhysRevLett.100.196804} {\bibfield  {journal}
  {\bibinfo  {journal} {Phys. Rev. Lett.}\ }\textbf {\bibinfo {volume} {100}},\
  \bibinfo {pages} {196804} (\bibinfo {year} {2008})}\BibitemShut {NoStop}%
\bibitem [{\citenamefont {Kiguchi}\ \emph {et~al.}(2008)\citenamefont
  {Kiguchi}, \citenamefont {Tal}, \citenamefont {Wohlthat}, \citenamefont
  {Pauly}, \citenamefont {Krieger}, \citenamefont {Djukic}, \citenamefont
  {Cuevas},\ and\ \citenamefont {van Ruitenbeek}}]{kiguchi_highly_2008}%
  \BibitemOpen
  \bibfield  {author} {\bibinfo {author} {\bibfnamefont {M.}~\bibnamefont
  {Kiguchi}}, \bibinfo {author} {\bibfnamefont {O.}~\bibnamefont {Tal}},
  \bibinfo {author} {\bibfnamefont {S.}~\bibnamefont {Wohlthat}}, \bibinfo
  {author} {\bibfnamefont {F.}~\bibnamefont {Pauly}}, \bibinfo {author}
  {\bibfnamefont {M.}~\bibnamefont {Krieger}}, \bibinfo {author} {\bibfnamefont
  {D.}~\bibnamefont {Djukic}}, \bibinfo {author} {\bibfnamefont {J.~C.}\
  \bibnamefont {Cuevas}}, \ and\ \bibinfo {author} {\bibfnamefont {J.~M.}\
  \bibnamefont {van Ruitenbeek}},\ }\href {\doibase
  10.1103/PhysRevLett.101.046801} {\bibfield  {journal} {\bibinfo  {journal}
  {Phys. Rev. Lett.}\ }\textbf {\bibinfo {volume} {101}},\ \bibinfo {pages}
  {046801} (\bibinfo {year} {2008})}\BibitemShut {NoStop}%
\bibitem [{\citenamefont {Song}\ \emph {et~al.}(2009)\citenamefont {Song},
  \citenamefont {Kim}, \citenamefont {Jang}, \citenamefont {Jeong},
  \citenamefont {Reed},\ and\ \citenamefont {Lee}}]{song_observation_2009}%
  \BibitemOpen
  \bibfield  {author} {\bibinfo {author} {\bibfnamefont {H.}~\bibnamefont
  {Song}}, \bibinfo {author} {\bibfnamefont {Y.}~\bibnamefont {Kim}}, \bibinfo
  {author} {\bibfnamefont {Y.~H.}\ \bibnamefont {Jang}}, \bibinfo {author}
  {\bibfnamefont {H.}~\bibnamefont {Jeong}}, \bibinfo {author} {\bibfnamefont
  {M.~A.}\ \bibnamefont {Reed}}, \ and\ \bibinfo {author} {\bibfnamefont
  {T.}~\bibnamefont {Lee}},\ }\href {\doibase 10.1038/nature08639} {\bibfield
  {journal} {\bibinfo  {journal} {Nature (London)}\ }\textbf {\bibinfo {volume} {462}},\
  \bibinfo {pages} {1039} (\bibinfo {year} {2009})}\BibitemShut {NoStop}%
\bibitem [{\citenamefont {Frederiksen}\ \emph {et~al.}(2007)\citenamefont
  {Frederiksen}, \citenamefont {Paulsson}, \citenamefont {Brandbyge},\ and\
  \citenamefont {Jauho}}]{frederiksen_inelastic_2007}%
  \BibitemOpen
  \bibfield  {author} {\bibinfo {author} {\bibfnamefont {T.}~\bibnamefont
  {Frederiksen}}, \bibinfo {author} {\bibfnamefont {M.}~\bibnamefont
  {Paulsson}}, \bibinfo {author} {\bibfnamefont {M.}~\bibnamefont {Brandbyge}},
  \ and\ \bibinfo {author} {\bibfnamefont {A.-P.}\ \bibnamefont {Jauho}},\
  }\href {\doibase 10.1103/PhysRevB.75.205413} {\bibfield  {journal} {\bibinfo
  {journal} {Phys. Rev. B}\ }\textbf {\bibinfo {volume} {75}},\ \bibinfo
  {pages} {205413} (\bibinfo {year} {2007})}\BibitemShut {NoStop}%
\bibitem [{\citenamefont {Ryndyk}\ \emph {et~al.}(2009)\citenamefont {Ryndyk},
  \citenamefont {Guti{\'e}rrez}, \citenamefont {Song},\ and\ \citenamefont
  {Cuniberti}}]{cuniberti_green_2009}%
  \BibitemOpen
  \bibfield  {author} {\bibinfo {author} {\bibfnamefont {D.~A.}\ \bibnamefont
  {Ryndyk}}, \bibinfo {author} {\bibfnamefont {R.}~\bibnamefont
  {Guti{\'e}rrez}}, \bibinfo {author} {\bibfnamefont {B.}~\bibnamefont {Song}},
  \ and\ \bibinfo {author} {\bibfnamefont {G.}~\bibnamefont {Cuniberti}},\ }in\
  \href {http://www.springerlink.com/content/n775121182u81552/abstract/} {\emph
  {\bibinfo {booktitle} {Energy Transfer Dynamics in Biomaterial Systems}}},\
  \bibinfo {series} {Springer Series in Chemical Physics}, Vol.~\bibinfo
  {volume} {93},\ \bibinfo {editor} {edited by\ \bibinfo {editor}
  {\bibfnamefont {I.}~\bibnamefont {Burghardt}}, \bibinfo {editor}
  {\bibfnamefont {V.}~\bibnamefont {May}}, \bibinfo {editor} {\bibfnamefont
  {D.~A.}\ \bibnamefont {Micha}}, \bibinfo {editor} {\bibfnamefont {E.~R.}\
  \bibnamefont {Bittner}}, \bibinfo {editor} {\bibfnamefont {A.~W.}\
  \bibnamefont {Castleman}}, \bibinfo {editor} {\bibfnamefont {J.~P.}\
  \bibnamefont {Toennies}}, \bibinfo {editor} {\bibfnamefont {K.}~\bibnamefont
  {Yamanouchi}}, \ and\ \bibinfo {editor} {\bibfnamefont {W.}~\bibnamefont
  {Zinth}}}\ (\bibinfo  {publisher} {Springer, Berlin/Heidelberg},\ \bibinfo
  {year} {2009})\ p.\ \bibinfo {pages} {213}\BibitemShut {NoStop}%
\bibitem [{\citenamefont {Lee}\ \emph {et~al.}(2009)\citenamefont {Lee},
  \citenamefont {Jean},\ and\ \citenamefont {Sanvito}}]{lee_exploring_2009}%
  \BibitemOpen
  \bibfield  {author} {\bibinfo {author} {\bibfnamefont {W.}~\bibnamefont
  {Lee}}, \bibinfo {author} {\bibfnamefont {N.}~\bibnamefont {Jean}}, \ and\
  \bibinfo {author} {\bibfnamefont {S.}~\bibnamefont {Sanvito}},\ }\href
  {\doibase 10.1103/PhysRevB.79.085120} {\bibfield  {journal} {\bibinfo
  {journal} {Phys. Rev. B}\ }\textbf {\bibinfo {volume} {79}},\ \bibinfo
  {pages} {085120} (\bibinfo {year} {2009})}\BibitemShut {NoStop}%
\bibitem [{\citenamefont {Knap}\ \emph {et~al.}(2011)\citenamefont {Knap},
  \citenamefont {von~der Linden},\ and\ \citenamefont
  {Arrigoni}}]{knap_noneq_2011}%
  \BibitemOpen
  \bibfield  {author} {\bibinfo {author} {\bibfnamefont {M.}~\bibnamefont
  {Knap}}, \bibinfo {author} {\bibfnamefont {W.}~\bibnamefont {von~der
  Linden}}, \ and\ \bibinfo {author} {\bibfnamefont {E.}~\bibnamefont
  {Arrigoni}},\ }\href {\doibase 10.1103/PhysRevB.84.115145} {\bibfield
  {journal} {\bibinfo  {journal} {Phys. Rev. B}\ }\textbf {\bibinfo {volume}
  {84}},\ \bibinfo {pages} {115145} (\bibinfo {year} {2011})}\BibitemShut
  {NoStop}%
\bibitem [{\citenamefont {Nuss}\ \emph {et~al.}(2012)\citenamefont {Nuss},
  \citenamefont {Heil}, \citenamefont {Ganahl}, \citenamefont {Knap},
  \citenamefont {Evertz}, \citenamefont {Arrigoni},\ and\ \citenamefont
  {von~der Linden}}]{nuss_steady-state_2012}%
  \BibitemOpen
  \bibfield  {author} {\bibinfo {author} {\bibfnamefont {M.}~\bibnamefont
  {Nuss}}, \bibinfo {author} {\bibfnamefont {C.}~\bibnamefont {Heil}}, \bibinfo
  {author} {\bibfnamefont {M.}~\bibnamefont {Ganahl}}, \bibinfo {author}
  {\bibfnamefont {M.}~\bibnamefont {Knap}}, \bibinfo {author} {\bibfnamefont
  {H.~G.}\ \bibnamefont {Evertz}}, \bibinfo {author} {\bibfnamefont
  {E.}~\bibnamefont {Arrigoni}}, \ and\ \bibinfo {author} {\bibfnamefont
  {W.}~\bibnamefont {von~der Linden}},\ }\href {\doibase 10.1103/PhysRevB.86.245119}
  {\bibfield  {journal} {\bibinfo  {journal} {{Phys. Rev. B}}\ } \textbf {\bibinfo {volume}
  {86}},\ \bibinfo {pages} {245119} (\bibinfo
  {year} {2012})}\BibitemShut {NoStop}%
\bibitem [{\citenamefont {Heeger}\ \emph {et~al.}(1988)\citenamefont {Heeger},
  \citenamefont {Kivelson}, \citenamefont {Schrieffer},\ and\ \citenamefont
  {Su}}]{heeger_solitons_1988}%
  \BibitemOpen
  \bibfield  {author} {\bibinfo {author} {\bibfnamefont {A.~J.}\ \bibnamefont
  {Heeger}}, \bibinfo {author} {\bibfnamefont {S.}~\bibnamefont {Kivelson}},
  \bibinfo {author} {\bibfnamefont {J.~R.}\ \bibnamefont {Schrieffer}}, \ and\
  \bibinfo {author} {\bibfnamefont {W.~P.}\ \bibnamefont {Su}},\ }\href
  {\doibase 10.1103/RevModPhys.60.781} {\bibfield  {journal} {\bibinfo
  {journal} {Rev. Mod. Phys.}\ }\textbf {\bibinfo {volume} {60}},\ \bibinfo
  {pages} {781} (\bibinfo {year} {1988})}\BibitemShut {NoStop}%
\bibitem [{\citenamefont {S{\'e}n{\'e}chal}\ \emph {et~al.}(2000)\citenamefont
  {S{\'e}n{\'e}chal}, \citenamefont {Perez},\ and\ \citenamefont
  {{Pioro-Ladri{\`e}re}}}]{snchal_spectral_2000}%
  \BibitemOpen
  \bibfield  {author} {\bibinfo {author} {\bibfnamefont {D.}~\bibnamefont
  {S{\'e}n{\'e}chal}}, \bibinfo {author} {\bibfnamefont {D.}~\bibnamefont
  {Perez}}, \ and\ \bibinfo {author} {\bibfnamefont {M.}~\bibnamefont
  {{Pioro-Ladri{\`e}re}}},\ }\href {\doibase 10.1103/PhysRevLett.84.522}
  {\bibfield  {journal} {\bibinfo  {journal} {Phys. Rev. Lett.}\ }\textbf
  {\bibinfo {volume} {84}},\ \bibinfo {pages} {522} (\bibinfo {year}
  {2000})}\BibitemShut {NoStop}%
\bibitem [{\citenamefont {Meir}\ and\ \citenamefont
  {Wingreen}(1992)}]{me.wi.92}%
  \BibitemOpen
  \bibfield  {author} {\bibinfo {author} {\bibfnamefont {Y.}~\bibnamefont
  {Meir}}\ and\ \bibinfo {author} {\bibfnamefont {N.~S.}\ \bibnamefont
  {Wingreen}},\ }\href {\doibase 10.1103/PhysRevLett.68.2512} {\bibfield
  {journal} {\bibinfo  {journal} {Phys. Rev. Lett.}\ }\textbf {\bibinfo
  {volume} {68}},\ \bibinfo {pages} {2512} (\bibinfo {year}
  {1992})}\BibitemShut {NoStop}%
\bibitem [{\citenamefont {Landauer}(1970)}]{landauer_electrical_1970}%
  \BibitemOpen
\bibfield  {journal} {  }\bibfield  {author} {\bibinfo {author} {\bibfnamefont
  {R.}~\bibnamefont {Landauer}},\ }\href {\doibase 10.1080/14786437008238472}
  {\bibfield  {journal} {\bibinfo  {journal} {Philos. Mag.}\ }\textbf {\bibinfo
  {volume} {21}},\ \bibinfo {pages} {863} (\bibinfo {year} {1970})}\BibitemShut
  {NoStop}%
\bibitem [{\citenamefont {B{\"u}ttiker}(1986)}]{buttiker_four-terminal_1986}%
  \BibitemOpen
  \bibfield  {author} {\bibinfo {author} {\bibfnamefont {M.}~\bibnamefont
  {B{\"u}ttiker}},\ }\href {\doibase 10.1103/PhysRevLett.57.1761} {\bibfield
  {journal} {\bibinfo  {journal} {Phys. Rev. Lett.}\ }\textbf {\bibinfo
  {volume} {57}},\ \bibinfo {pages} {1761} (\bibinfo {year}
  {1986})}\BibitemShut {NoStop}%
\bibitem [{\citenamefont {Zhu}\ and\ \citenamefont
  {Balatsky}(2003)}]{zhu_theory_2003}%
  \BibitemOpen
  \bibfield  {author} {\bibinfo {author} {\bibfnamefont {J.-X.}\ \bibnamefont
  {Zhu}}\ and\ \bibinfo {author} {\bibfnamefont {A.~V.}\ \bibnamefont
  {Balatsky}},\ }\href {\doibase 10.1103/PhysRevB.67.165326} {\bibfield
  {journal} {\bibinfo  {journal} {Phys. Rev. B}\ }\textbf {\bibinfo {volume}
  {67}},\ \bibinfo {pages} {165326} (\bibinfo {year} {2003})}\BibitemShut
  {NoStop}%
\bibitem [{\citenamefont {Romeike}\ \emph {et~al.}(2006)\citenamefont
  {Romeike}, \citenamefont {Wegewijs}, \citenamefont {Hofstetter},\ and\
  \citenamefont {Schoeller}}]{romeike_quantum_2006}%
  \BibitemOpen
  \bibfield  {author} {\bibinfo {author} {\bibfnamefont {C.}~\bibnamefont
  {Romeike}}, \bibinfo {author} {\bibfnamefont {M.~R.}\ \bibnamefont
  {Wegewijs}}, \bibinfo {author} {\bibfnamefont {W.}~\bibnamefont
  {Hofstetter}}, \ and\ \bibinfo {author} {\bibfnamefont {H.}~\bibnamefont
  {Schoeller}},\ }\href {\doibase 10.1103/PhysRevLett.96.196601} {\bibfield
  {journal} {\bibinfo  {journal} {Phys. Rev. Lett.}\ }\textbf {\bibinfo
  {volume} {96}},\ \bibinfo {pages} {196601} (\bibinfo {year}
  {2006})}\BibitemShut {NoStop}%
\bibitem [{\citenamefont {Rai}\ \emph {et~al.}(2010)\citenamefont {Rai},
  \citenamefont {Hod},\ and\ \citenamefont {Nitzan}}]{rai_circular_2010}%
  \BibitemOpen
  \bibfield  {author} {\bibinfo {author} {\bibfnamefont {D.}~\bibnamefont
  {Rai}}, \bibinfo {author} {\bibfnamefont {O.}~\bibnamefont {Hod}}, \ and\
  \bibinfo {author} {\bibfnamefont {A.}~\bibnamefont {Nitzan}},\ }\href
  {\doibase 10.1021/jp105030d} {\bibfield  {journal} {\bibinfo  {journal} {J.
  Phys. Chem. C}\ }\textbf {\bibinfo {volume} {114}},\ \bibinfo {pages} {20583}
  (\bibinfo {year} {2010})}\BibitemShut {NoStop}%
\end{thebibliography}

%

\end{document}